\documentclass[prb,twocolumn,showpacs]{revtex4}
\usepackage[dvips]{graphicx}
\usepackage{latexsym}
\usepackage{amsmath}
\usepackage{amsfonts}
\usepackage{amssymb}
\usepackage{bm}
\usepackage{color}
\begin{document}
\newcommand{\fig}[2]{\includegraphics[width=#1]{#2}}
\newcommand{{\vhf}}{\chi^\text{v}_f}
\newcommand{{\vhd}}{\chi^\text{v}_d}
\newcommand{{\vpd}}{\Delta^\text{v}_d}
\newcommand{{\ved}}{\epsilon^\text{v}_d}
\newcommand{{\vved}}{\varepsilon^\text{v}_d}
\newcommand{{\bk}}{{\bf k}}
\newcommand{{\br}}{{\vec r}}
\newcommand{{\bq}}{{\bf q}}
\newcommand{{\te}}{{\tilde{e}}}
\newcommand{{\td}}{{\tilde{d}}}
\newcommand{{\tp}}{{\tilde{p}}}
\newcommand{{\uu}}{{\tilde{u}}}
\newcommand{{\uv}}{{\tilde{v}}}
\newcommand{\pprl}{Phys. Rev. Lett. \ }
\newcommand{\pprb}{Phys. Rev. {B}}

\title{Doublon-holon binding, Mott transition, and fractionalized antiferromagnet \\
in the Hubbard model}
\author{Sen Zhou,$^{1,2}$ Yupeng Wang,$^3$ and Ziqiang Wang$^2$}
\affiliation{$^1$ State Key Laboratory of  Theoretical Physics, Institute of Theoretical Physics, Chinese Academy of Sciences, Beijing 100190, China}
\affiliation{$^2$ Department of Physics, Boston College, Chestnut Hill, MA 02467, USA}
\affiliation{$^3$ The Institute of Physics, Chinese Academy of Sciences, Beijing 100190, China}
\date{\today}

\begin{abstract}

We argue that the binding between doubly occupied (doublon) and empty (holon) sites governs the incoherent excitations and plays a key role in the Mott transition in strongly correlated Mott-Hubbard systems.
We construct a new saddle point solution with doublon-holon binding in the Kotliar-Ruckenstein slave-boson functional integral formulation of the Hubbard model.
On the half-filled honeycomb lattice and square lattice, the ground state is found to exhibit a continuous transition from the paramagnetic semimetal/metal to an antiferromagnetic ordered Slater insulator with coherent quasiparticles at $U_{c1}$, followed by a Mott transition into an electron-fractionalized AF$^*$ phase without coherent excitations at $U_{c2}$. Such a phase structure appears generic of bipartite lattices without frustration. We show that doublon-holon binding unites the three important ideas of strong correlation: the coherent quasiparticles, the incoherent Hubbard bands, and the deconfined Mott insulator.

\typeout{polish abstract}
\end{abstract}

\pacs{71.10.-w, 71.27.+a, 71.10.Fd, 71.30.+h}

\maketitle

\section{Introduction}
The fundamental theoretical challenge of the strong correlation problem is the description of both the low energy coherent quasiparticles (QPs) and the higher energy incoherent excitations, and the spectral weight transfer from coherent to incoherent excitations with increasing correlation strength.
Two very important ideas, the emergence of two broad incoherent features known as the Hubbard bands and the existence of renormalized QPs with a Luttinger Fermi surface (FS) were advanced by Hubbard \cite{hubbard}, and Brinkman and Rice \cite{brinkman}, respectively.
Unfortunately, the Hubbard equation of motion scheme that produces the incoherent spectrum fails to produce QPs correctly and violates Luttinger's theorem \cite{luttinger}; whereas the approaches based on the Brinkman-Rice-Gutzwiller wave functions \cite{gutzwiller} capture a Luttinger FS of QPs, but find serious difficulties in constructing variational excited states to account for the incoherent excitations.
Faced with this enigma, numerical approaches such as exact diagonalization, quantum Monte Carlo (QMC), and the dynamical mean field theory (DMFT) \cite{dmft} have played a key role in recent studies of the strong correlation problem.

In this paper, we develop new analytical insights and construct a unified theory for both the coherent and incoherent excitations as well as the magnetic and the Mott transition. Our focus will be the half-filled single-band Hubbard model on bipartite lattices without frustration. As specific examples, we study the square lattice and the honeycomb lattice especially in view of the recent debate over the possible emergence of a gapped spin liquid (SL) phase on the honeycomb lattice \cite{meng, wliu, liebsch, zylu, jxli, hur12, sorella,  seki, hassan13, liebsch2}. With only on-site interactions, the Hilbert space of the Hubbard model is a product of the local Hilbert space on a single-site that consists of empty (holon), doubly occupied (doublon), and singly occupied states. The Brinkman-Rice-Gutzwiller approach amounts to a metallic state where the holon, denoted as a boson $e$, and the doublon, as $d$, condense fully with macroscopic phase coherence, as can be obtained by the Gutzwiller approximation \cite{gwa-finiteU} or the slave boson mean-field theory \cite{kr}.
The metal-insulator transition is thus forced to follow a route where the \emph{density} of doublons and holons vanish together with the condensates:
$n_d=n_e=\langle d\rangle=\langle e\rangle=0$ such that there is
\emph{exactly} one electron per site.
As a result, single-particle motion, coherent or incoherent, is completely prohibited.
This so-called Brinkman-Rice (BR) transition is different from the Mott transition induced by the complete transfer of the coherent QP weight into the incoherent background, {\em i.e.} the depletion of the condensate while keeping the doublon/holon (D/H) \emph{density} finite in the Mott insulator.

We will show that the crucial physics uniting the disparate ideas of Hubbard and BR is the binding between doublon and holon:
$\langle d_i e_j\rangle\ne0$.
In the Mott insulator at large $U$, although the D/H condensate vanishes ($\langle d \rangle =\langle e\rangle =d_0=0$) together with the disappearance of the coherent QP, the D/H \emph{density} remains nonzero ($n_d=n_e\ne0$).
The motion of the QP is thus possible by breaking the doublon-holon (D-H) pairs, giving rise to the higher-energy incoherent excitations.
With decreasing $U$, the D/H density increases and the D-H binding energy decreases.
At a critical $U_c$, the D-H excitation gap closes and the D/H
single-particle condensate develops, marking the onset of the Mott transition.
On the metallic side of the Mott transition, D-H binding continues to play an important role since an added electron can propagate either as a coherent QP via the D/H condensate or incoherently via the unbinding of the D-H pairs.

The idea that D-H binding plays an important role in Mott-Hubbard systems was introduced by Kaplan, Horch, and Fulde \cite{khf82} and studied in the context of improved variational Gutzwiller wave functions \cite{yokoyama, capello05}.
The difficulty in constructing the appropriate variational wave functions for excitations has prevented further advances along these lines.
More recently, field theory approaches involving the binding of charge 2$e$ doublons with fermionic quasiparticles \cite{phillips09} as well as combining the bosons with fermions to form co-fermions \cite{imada11} have been put forth within the context of doped Mott insulators.

\begin{figure}
\begin{center}
\fig{3.4in}{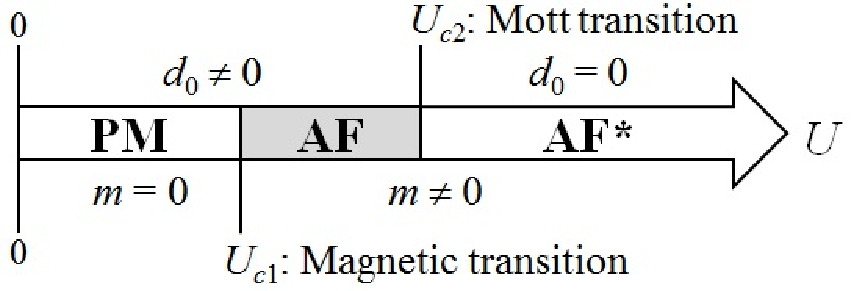} \caption{Schematic phase diagram obtained for the half-filled single-band Hubbard model with D-H binding on the honeycomb lattice and square lattice.
The AF order $m$ is developed after the magnetic transition $U_{c1}$, and the single-boson condensate of doublon $d_0$ disappears at the Mott transition $U_{c2}$.}
\label{diagram}
\end{center}
\end{figure}

In this work, we will show that the physical picture presented above can be realized in the slave-boson functional integral formulation of the Hubbard model introduced by Kotliar and Ruckenstein (KR) \cite{kr} by constructing new saddle point solutions that include the D-H binding.
This approach also offers a treatment of the magnetism at half-filling that compares well to QMC simulations \cite{werner} and has the added advantage of allowing the study of excitations and finite temperature properties \cite{otherslaveboson}.
As concrete examples, we studied the D-H binding in the half-filled Hubbard model on the honeycomb and the square lattice at zero temperature and obtained the phase diagram shown schematically in Fig.~1.
On the honeycomb lattice, a continuous transition from the semimetal (SM) to an antiferromagnetic (AF) ordered insulator takes place at a critical Coulomb repulsion $U_{c1} \simeq 3.4t$, suggesting that the gapped SL phase proposed by Meng et. al. \cite{meng} may correspond to an AF ordered phase in the thermodynamic limit.
Sorella et. al. \cite{sorella} recently extended the QMC and the finite size scaling analysis to much larger system sizes and discovered that the signature of the gapped SL disappears and is replaced by that of a continuous SM to AF order transition at $U\simeq3.8t$, in qualitative agreement with our results.
Remarkably, we found a second quantum phase transition at a critical $U_{c2} \simeq 5.7t$ beyond which the D/H single-particle condensate vanishes ($d_0=0$) amid a finite density of doublons bound to holons.
For $U>U_{c2}$, a new AF phase without coherent QP excitations, termed as AF$^*$ in Fig.~1, emerges where the electrons are fractionalized and the elementary excitations do not carry the quantum numbers of an electron.
We obtained a similar phase diagram on the square lattice; the transition to the Slater AF state happens for infinitesimal $U$ (i.e. $U_{c1}=0$) due to the perfect nesting of the Fermi surface on the square lattice while the transition to the AF* phase takes place at $U_{c2}\simeq6.8t$.

The rest of the paper is organized as follows.
Section II describes the slave-boson path integral formulation of the Hubbard model, the KR saddle point solution on the honeycomb and the square lattice, and the BR metal-insulator transition.
In section III, we introduce the slave boson intersite correlations into the functional integral and construct the new saddle-point solution that includes the effects of D-H binding.
The Mott transition and the spectral weight transfer between coherent and incoherent excitations will be studied to obtain the phase diagrams of the Hubbard model on the two half-filled bipartite lattices. We describe the transitions between the paramagnetic metal/semimetal, Slator AF insulator, and AF$^*$ phases, elucidate the properties of the electron-fractionalized AF$^*$ phase, and develop further insights into the nature of the incoherent excitations in Mott-Hubbard systems.
Section IV contains the summary and conclusions.

\section{Hubbard model and slave boson functional integral representation}

We start with the Hubbard model with nearest neighbor hopping $t$ and on-site Coulomb repulsion $U$,
\begin{equation}
\hat{H}=-t\sum_{\langle i,j\rangle,\sigma} \big( c^\dagger_{i\sigma} c_{j\sigma} +h.c. \big) +U\sum_i \hat{n}_{i\uparrow} \hat{n}_{i\downarrow},
\end{equation}
where $c^\dagger_{i\sigma}$ creates an electron with spin $\sigma$ on site $i$, and $\hat{n}_{i\sigma} =c^\dagger_{i\sigma} c_{i\sigma}$ is the density operator.
In the KR formulation \cite{kr}, the electron operator is written as
\begin{equation}
c_{i\sigma}= {\hat z}_{i\sigma} f_{i\sigma}, \qquad  {\hat
z}_{i\sigma}={\hat L}_{i\sigma} (e_i^\dagger p_{i\sigma}
+p_{i{\bar\sigma}}^\dagger d_i) {\hat R}_{i\sigma},
\end{equation}
where the boson operators describe the holon ($e_i$), doublon ($d_i$), and singly-occupied ($p_{i\sigma}$) sites, and $f_{i\sigma}$ is a fermion operator.
The operators ${\hat L}_\sigma$ and ${\hat R}_\sigma$ are diagonal
with unit eigenvalues in the (empty, $\bar \sigma$) and the ($\sigma$, doubly-occupied) subspaces, respectively \cite{lavagna},
\begin{equation}
{\hat L}_{i\sigma}= (1-d_i^\dagger d_i-p_{i\sigma}^\dagger
p_{i\sigma})^\alpha, \hspace{0.1cm} {\hat R}_{i\sigma}= (1-e_i^\dagger e_i -p_{i\bar\sigma}^\dagger p_{i\bar\sigma})^\beta, \nonumber
\end{equation}
where $\alpha$ and $\beta$ can take any value. The Hubbard model Hamiltonian is thus given by,
\begin{align}
\hat{H}_\text{KR}=&-t\sum_{\langle i,j\rangle,\sigma} \big( {\hat z}^\dagger_{i\sigma} {\hat z}_{j\sigma} f^\dagger_{i\sigma} f_{j\sigma} +h.c.\big) +U\sum_i d^\dagger_i d_i.
\label{Hsb}
\end{align}
The partition function is a coherent state path integral over the quantum fields \cite{normalorder}:
\begin{equation}
Z=\int \mathcal{D} [f,f^\dagger] \mathcal{D}[e,e^\dagger]
\mathcal{D}[p,p^\dagger] \mathcal{D}[d,d^\dagger]
\mathcal{D}[\lambda,\lambda_\sigma] e^{ -\int^\beta_0 \mathcal{L} d\tau},
\label{partition1}
\end{equation}
where the Lagrangian is given by
\begin{align}
\mathcal{L}=& \sum_i ( e^\dagger_i {\partial_\tau} e_i + d^\dagger_i {\partial_\tau} d_i) +\sum_{i,\sigma} ( p^\dagger_{i\sigma} {\partial_\tau} p_{i\sigma} + f^\dagger_{i\sigma} {\partial_\tau} f_{i\sigma}) \nonumber \\
+&\hat{H}_\text{KR} +i\sum_i \lambda_i \hat{Q}_i +i\sum_{i,\sigma} \lambda_{i\sigma} \hat{Q}_{i\sigma} -\mu\sum_{i\sigma} f^\dagger_{i\sigma}f_{i\sigma},
\label{lagrangian1}
\end{align}
where $\mu$ is the chemical potential and  $\lambda_i$ and
$\lambda_{i\sigma}$ are the Lagrange multipliers introduced to enforce the local constraints for the completeness of the Hilbert space:
\begin{align}
\hat{Q}_{i} &=e^\dagger_i e_i +\sum_\sigma p^\dagger_{i\sigma} p_{i\sigma} + d^\dagger_i d_i - 1 =0, \label{constraint:1}
\end{align}
and the equivalence between the fermion and boson representations of the spin-dependent density:
\begin{align}
\hat{Q}_{i\sigma} &= f^\dagger_{i\sigma} f_{i\sigma} -p^\dagger_{i\sigma}
p_{i\sigma} - d^\dagger_i d_i =0. \label{constraint:2}
\end{align}
The KR saddle point corresponds to condensing all the boson fields uniformly with their values determined by minimizing the action \cite{kr}.
KR found that for $\alpha=\beta=-1/2$, the saddle point solution recovers the Gutzwiller approximation \cite{kr}.

\begin{figure}
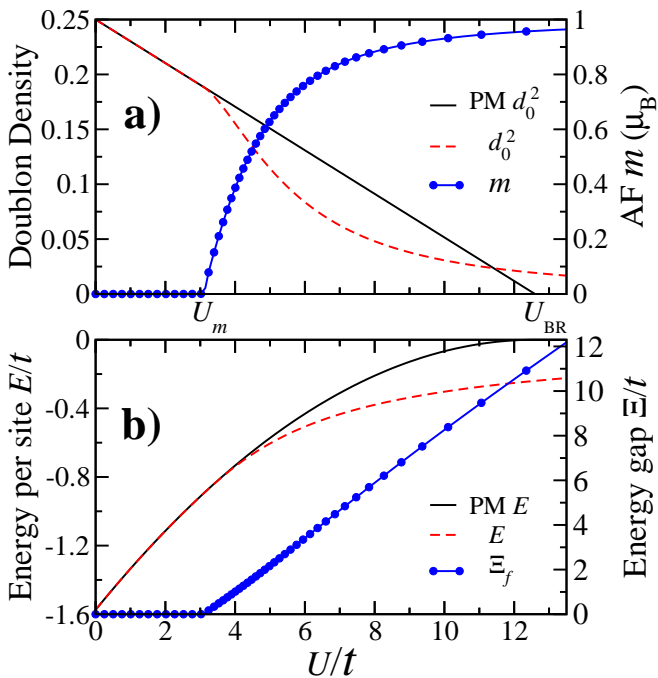

\begin{center}
\fig{3.4in}{honeycombGW.eps} \caption{(color online). KR saddle-point solutions of the Hubbard model on the half-filled honeycomb lattice.
The Hubbard $U$ dependence of (a) doublon condensate density $d^2_0$, staggered magnetization $m$, and (b) energy per site $E$ and the single-particle energy gap in the fermion sector $\Xi_f$.
The corresponding results in the restricted PM phase are also shown.} \label{honeycombGW}
\end{center}
\end{figure}

\begin{figure}
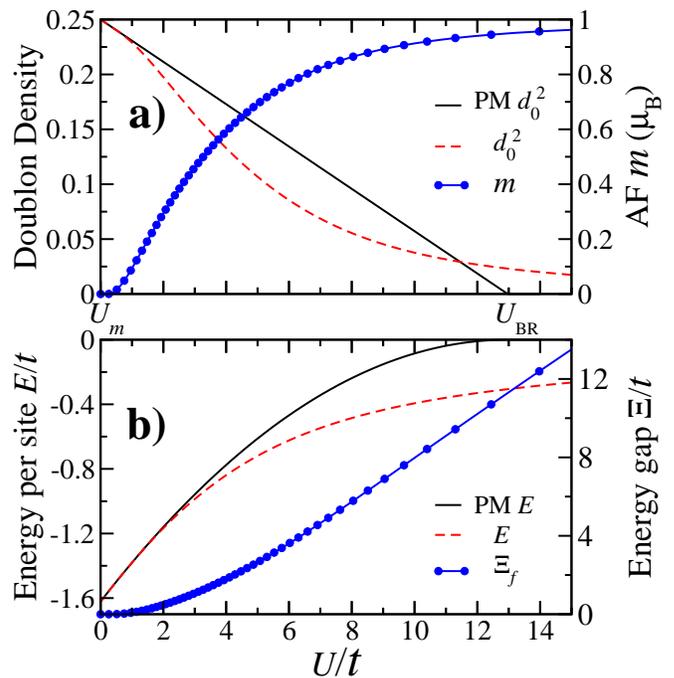

\begin{center}
\fig{3.4in}{squareGW.eps} \caption{(color online). KR saddle-point solutions of the Hubbard model on the half-filled square lattice.
Labels and notations follow those in Fig. \ref{honeycombGW}.} \label{squareGW}
\end{center}
\end{figure}

The KR saddle point solutions on the half-filled honeycomb lattice \cite{GWhoneycomb} are summarized in Fig.~\ref{honeycombGW}.
Restricting to the paramagnetic (PM) phase, the doublon density $d^2_0$ decreases linearly from 1/4 at $U=0$ and vanishes at the BR metal-insulator transition $U_{BR}\simeq 12.6 t$.
When magnetism is allowed, a SM to an AF insulator transition arises at $U_m\simeq 3.1t$.
The D/H condensate remains nonzero for all finite $U$ and the AF phase is a Slater insulator with coherent quasiparticle excitations.
The results on the half-filled square lattice are shown in Fig.~\ref{squareGW}.
The BR metal-insulator takes place at $U_{BR}\simeq 13t$.
When magnetism is allowed, the PM metal is unstable with respect to the Slator AF insulator for any nonzero $U$ owning to the perfect nesting of the Fermi surface; the AF ordered moment develops exponentially at $U_m=0$.

\section{Boson inter-site correlations and new saddle-point solutions with
doublon-holon binding}

The KR saddle-point solution, i.e. the Gutzwiller approximation, ignores all inter-site correlations and captures only the coherent QP single-particle excitations. Indeed, it has been shown \cite{GWinf-d} that in the limit of infinite dimensions (infinite-$d$), the latter becomes an exact solution of the variational Gutzwiller wave function approach. Our strategy for going beyond the Gutzwiller approximation described by the KR saddle point is to build in explicitly the inter-site correlations and boson dynamics in the functional integral and construct a new saddle point solution that includes D-H binding.

\subsection{Path integral including boson inter-site correlations}

Introducing the operators for the D-H pairing $\hat{\Delta}_{ij} =d_ie_j$, and the D/H hopping $\hat{\chi}^d_{ij} =d^\dagger_i d_j$, $\hat{\chi}^e_{ij} =e^\dagger_i e_j$ on the nearest-neighbor bonds, as well as the density operators $\hat{n}^d_i =d^\dagger_i d_i$, $\hat{n}^e_i =e^\dagger_i e_i$ on each site, we can
rewrite the bosonic part in the hopping term in Eq.~(\ref{Hsb}) as
\begin{align}
{\hat z}_{i\sigma}^\dagger {\hat z}_{j\sigma} &={\hat Y}_{ij,\sigma}^{-1/2}
[(\hat{\chi}^e_{ij})^\dagger p^\dagger_{i\sigma} p_{j\sigma} +
\hat{\chi}^d_{ij} p_{i\bar\sigma}  p^\dagger_{j\bar\sigma} \label{zz} \\
&\hspace{1.2cm} +\hat{\Delta}_{ji} p^\dagger_{i\sigma}
p^\dagger_{j\bar\sigma} + \hat{\Delta}_{ij}^\dagger p_{i\bar\sigma} p_{j\sigma}] {\hat Y}_{ji,\sigma}^{-1/2}, \nonumber
\end{align}
where
\begin{align}
\hat{Y}_{ij,\sigma}&= \hat{R}^{-2}_{i\sigma} \hat{L}^{-2}_{j\sigma}= \big[(1-
p^\dagger_ {i\bar\sigma} p_{i\bar\sigma}) (1- p^\dagger_ {j\sigma}
p_{j\sigma}) \label{Yijs} \\
&\hspace{0.35cm} -\hat{n}^e_{i} (1- p^\dagger_ {j\sigma}
p_{j\sigma})-\hat{n}^d_{j} (1- p^\dagger_ {i\bar\sigma} p_{i\bar\sigma})
+|\hat{\Delta}_{ji}|^2 \big]. \nonumber
\end{align}
Note that, due to the normal ordering of the square roots, the
expression for $\hat{Y}_{ij,\sigma}$ involves explicitly the D-H pairing, but not the H/D hopping operators.
The obvious challenge is how to build these correlations into the
calculation of the partition function.
Since they enter through the rather formidable factor
$\hat{z}^\dagger_{i\sigma} \hat{z}_{j\sigma}$, the usual procedure of
introducing the corresponding correlation fields ($\Delta_{ij}$,
$\chi_{i,j}^e$, $\chi_{ij}^d$, $n_i^d$, $n_i^e$) via Hubbard-Stratonovich transformations in the path-integral does not work here.
We found that the difficulty can be overcome by introducing in the
functional integral additional Lagrange multipliers in the corresponding channel, $\Delta^\text{v}_{ij}$, $\chi^{d,\text{v}}_{ij}$,
$\chi^{e,\text{v}}_{ij}$, $\epsilon^{d,\text{v}}_i$, and
$\epsilon^{e,\text{v}}_i$, such that the partition function becomes:
\begin{align}
Z=&\int \mathcal{D} [f,f^\dagger] \mathcal{D}[e,e^\dagger]
\mathcal{D}[p,p^\dagger] \mathcal{D}[d,d^\dagger]
\mathcal{D}[\lambda,\lambda_\sigma] \label{partition2} \\
&\mathcal{D}[\Delta,\chi^d,\chi^e,n^d,n^e] \mathcal{D}[\Delta^\text{v},
\chi^{d,\text{v}}, \chi^{e,\text{v}},
\epsilon^{d,\text{v}}\epsilon^{e,\text{v}}] e^{-\int_0^\beta d\tau {\cal L}},
\nonumber
\end{align}
with the Lagrangian
\begin{align}
{\cal L}&=\sum_i ( e^\dagger_i {\partial_\tau} e_i + d^\dagger_i
{\partial_\tau} d_i) +\sum_{i,\sigma} ( p^\dagger_{i\sigma}{\partial_\tau}
p_{i\sigma} + f^\dagger_{i\sigma} {\partial_\tau} f_{i\sigma}) \nonumber \\
&+ \hat{H}_\text{DH} +i\sum_i \lambda_i \hat{Q}_i +i\sum_{i,\sigma}
\lambda_{i\sigma} \hat{Q}_{i\sigma} -\mu\sum_{i\sigma} f^\dagger_{i\sigma}
f_{j\sigma}, \label{L2}
\end{align}
where $\hat{H}_\text{DH}$ is the effective D-H binding Hamiltonian
\begin{align}
\hat{H}_\text{DH}=
&-t\sum_{\langle i,j\rangle,\sigma} ( {z}_{i\sigma} {z}_{j\sigma}
f^\dagger_{i\sigma} f_{j\sigma} +h.c.) + U\sum_i d^\dagger_i d_i \nonumber \\
& -i\sum_{\langle i,j\rangle} \Big[ \chi^{d,\text{v}}_{ij} ( d^\dagger_i
d_j -\chi^d_{ij}) + \chi^{e,\text{v}}_{ij} (e^\dagger_i e_j -\chi^e_{ij} ) \nonumber \\
& \hspace{0.2cm}+ \Delta^{\text{v}}_{ij} (d_i e_j - \Delta_{ij}) + \Delta^{\text{v}}_{ji} (e_i d_j - \Delta_{ji})+h.c.\Big] \nonumber \\
&+i\sum_i \left[ \epsilon^{d,\text{v}}_i ( d^\dagger_i d_i - n^d_i) +\epsilon^{e,\text{v}}_i ( e^\dagger_i e_i - n^e_i) \right]. \label{HDH}
\end{align}
The factor ${z}_{i\sigma} {z}_{j\sigma}$ in Eq.~(\ref{HDH}) has the same form given in Eqs. (\ref{zz}) and (\ref{Yijs}) but with the bosonic operators replaced by their corresponding correlation fields ($\Delta_{ij},\chi_{i,j}^e, \chi_{ij}^d, n_i^d, n_i^e$).
A few remarks are in order.
(i) Eqs. (\ref{partition2}-\ref{HDH}) provide an exact representation of the Hubbard model;
carrying out formally the last two functional integrals
in Eq.~(\ref{partition2}) recovers the KR formulation given in Eqs. (\ref{Hsb}-\ref{lagrangian1}).
(ii) The intersite correlations of the $p_{i\sigma}$ bosons can be included in a similar manner.
For simplicity, we treat the latter as condensed fields in this paper since their densities (\textit{i.e.}, the density of single occupations) remain large at half-filling.
(iii) From the perspective of finding a saddle point solution of the action, the effective Hamiltonian $H_\text{DH}$ in Eq.~(\ref{HDH}) can be understood intuitively as a variational Hamiltonian describing the effects of intersite correlations of the doublons and holons, including that of D-H binding, where $\Delta^{\text{v}}_{ij}$, $\chi^{d,\text{v}}_{ij}$,
$\chi^{e,\text{v}}_{ij}$, $\epsilon^{d,\text{v}}_i$, and
$\epsilon^{e,\text{v}}_i$ are nothing but the variational
parameters to be self-consistently determined.

\subsection{Saddle point solutions with D-H binding}

We next discuss the D-H binding saddle point solutions of
the path integral in Eqs.~(\ref{partition2}-\ref{HDH}) which correspond to configurations of the quantum fields that minimize the action.
We consider here that translation invariant PM and the two-sublattice AF saddle point solutions on the half-filled bipartite (honeycomb and square) lattices with $2N$ sites.
The bond variables are taken to be real and isotropic, and symmetry requires
$\Delta_{\langle ij\rangle} =\Delta_d$, $\chi^d_{\langle ij\rangle} =\chi^e_{\langle ij\rangle} =\chi_d$, $n^d_i =n^e_i =n_d$, and
correspondingly, $i\Delta^\text{v}_{\langle ij\rangle} =\vpd$, $i\chi^{d, \text{v}} _{\langle ij\rangle} =i\chi^{e,\text{v}}_{\langle ij \rangle} =\vhd$, $i\epsilon^{d,\text{v}}_i =i\epsilon^{e,\text{v}}_i =\ved$.
Moreover $i\lambda_i =\lambda$, $i\lambda_{A\sigma} =
i\lambda_{B\bar\sigma} =\lambda_\sigma$, and $p_{A0\sigma} = p_{B0\bar\sigma} = p_{0\sigma}$, where $A$ and $B$ denote the two sublattices on the bipartite lattice.
Consequently, on the nearest neighbor bonds $\langle i,j\rangle$, the factor
\begin{equation}
t{z}_{i\sigma} {z}_{j\sigma} = tg \left[2p_{0\uparrow} p_{0\downarrow} \chi_d +(p^2_{0\uparrow} +p^2_{0\downarrow}) \Delta_d \right] \equiv \vhf,
\end{equation}
where $g= \prod_\sigma Y^{-1/2}_\sigma$, with
\begin{equation}
Y_\sigma = 1- 2n_d -2p^2_{0\sigma} +2p^2_{0\sigma} n_d +p^4_{0\sigma} +\Delta^2_d .
\end{equation}
As will be shown later, this expression ensures that the new saddle point solution recovers the noninteracting limit at $U=0$.
Substituting these quantities into Eq.~(\ref{HDH}), we obtain the saddle point Hamiltonian,
\begin{align}
\hat{H}_\text{DH}^\text{sp}&= \hat{H}_f + \hat{H}_d +4 N \zeta  (\vhd \chi_d + \vpd \Delta_d) \label{HDHsp}\\
&-2N(\ved+ 2\lambda) -4N\ved n_d +2N\sum_\sigma (\lambda- \lambda_\sigma) p^2_{0\sigma} \nonumber
\end{align}
where the coordination number $\zeta=3$ on the honeycomb lattice, and $\zeta=4$ on the square lattice.
The Hamiltonian $\hat{H}_f$ and $\hat{H}_d$ determines the energy spectra in the fermion and boson sectors, respectively.

\subsubsection{Fermion spectrum}

The fermion spectrum is given by, in terms of the wave vector $\bk$ defined on the reciprocal lattice,
\begin{equation}
\hat{H}_f = \sum_{\bk,\sigma} \left[ \begin{array}{c} f_{A\bk\sigma}
\\ f_{B\bk\sigma} \end{array} \right]^\dagger \left[
\begin{array}{cc} \lambda_\sigma -\mu & - \vhf \eta_\bk \\
-\vhf \eta^*_\bk & \lambda_{\bar\sigma} -\mu \end{array} \right]
\left[ \begin{array}{c} f_{A\bk\sigma} \\ f_{B\bk\sigma}
\end{array} \right], \label{Efermion}
\end{equation}
where $\eta_\bk$ is the dispersion due to the nearest neighbor hopping $t$, which takes the form of
\begin{equation}
\eta_\bk =\exp(ik_y) + 2\cos(\sqrt{3}k_x/2) \exp(-ik_y/2)
\nonumber
\end{equation}
on the honeycomb lattice and
\begin{equation}
\eta_\bk =2(\cos{k_x} +\cos{k_y})
\nonumber
\end{equation}
on the square lattice.
The sum over $\bk$ runs over the first Brillouin zone corresponding to a unit cell with two sites.
The particle-hole symmetry at half-filling requires
$\mu=U/2$ and $\lambda_\sigma =\mu -\sigma \varepsilon$, where
$\varepsilon= (\lambda_\uparrow -\lambda_\downarrow)/2$ becomes nonzero when the AF order develops.
The fermion dispersion is thus obtained by diagonalizing
Eq.~(\ref{Efermion}),
\begin{equation}
E^f_\pm(\bk) = \pm \sqrt{\varepsilon^2 + |\vhf \eta_\bk|^2}. \label{ekf}
\end{equation}
A gap of $\Xi_f = 2|\varepsilon|$ would open in the fermion spectrum in the presence of AF order.

On the square lattice, the sublattices $A$ and $B$ become equivalent in the PM phase where the fermion spectrum in Eq. (\ref{Efermion}) simplifies to
\begin{equation}
\hat{H}_f = -\vhf \sum_{\bk,\sigma} \eta_\bk f^\dagger_{\bk\sigma} f_{\bk\sigma}.
\end{equation}
Here, the sum over $\bk$ runs over the first Brillouin zone corresponding to a unit cell containing only one site.

\subsubsection{Boson spectrum}

The charged boson spectrum is governed by
\begin{equation}
\hat{H}_d = \sum_\bk \Psi^\dagger_\bk M_\bk \Psi_\bk, \quad
\Psi_\bk =\big[ d_{A\bk}, d_{B\bk},
e^\dagger_{B\bar\bk},e^\dagger_{A\bar\bk} \big]^T,
\nonumber
\end{equation}
where the boson Hamiltonian matrix
\begin{equation}
M_\bk =\left[ \begin{array}{cccc} \ved+\lambda & -\vhd \eta_\bk & -\vpd
\eta_\bk & 0 \\ -\vhd \eta^*_\bk & \ved+\lambda & 0 & -\vpd \eta^*_\bk \\
-\vpd \eta^*_\bk & 0 & \ved+\lambda & -\vhd \eta^*_\bk \\ 0 & -\vpd \eta_\bk
& -\vhd \eta_\bk & \ved+\lambda \end{array} \right]. \label{Eboson}
\end{equation}
Here the relations due to the particle-hole symmetry at half-filling have been applied.
Note that $M_\bk$ is independent of spin.
The boson dispersion is obtained by diagonalizing Eq.~(\ref{Eboson}) using the standard boson Bogoliubov transformation:
\begin{equation}
E^d_\pm (\bk) =\sqrt{\big( \ved +\lambda \pm |\vhd \eta_\bk| \big)^2 - |\vpd \eta_\bk|^2 }. \label{ekb}
\end{equation}
Each branch of the above dispersion is doubly degenerate.
The condition for a real physical dispersion requires that $\ved +\lambda \ge \zeta \left( |\vhd| + |\vpd| \right)$.
When the equality is satisfied, the boson spectrum is gapless and the bosons can condense into the zero energy state. Otherwise, an energy gap
\begin{equation}
\Xi_d = 2\sqrt{\left( \ved +\lambda - \zeta |\vhd|
\right)^2 - \left( \zeta |\vpd| \right)^2}
\nonumber
\end{equation}
develops in the boson spectrum and the doublon and holon condensate would be depleted.

In the PM phase on the square lattice, the sublattices $A$ and $B$ become equivalent and the boson Hamiltonian matrix in Eq. (\ref{Eboson}) simplifies to
\begin{equation}
\hat{H}_d = \sum_{\bk,\sigma} \left[ \begin{array}{c} d_\bk
\\ e^\dagger_{\bar\bk} \end{array} \right]^\dagger \left[
\begin{array}{cc} \ved+\lambda -\vhd \eta_\bk & - \vpd \eta_\bk \\
- \vpd \eta_\bk & \ved+\lambda -\vhd \eta_\bk \end{array} \right]
\left[ \begin{array}{c}  d_\bk \\ e^\dagger_{\bar\bk} \end{array} \right].\label{Eboson2}
\end{equation}
This results in a doubly degenerate boson dispersion relation
\begin{equation}
E^d (\bk) =\sqrt{\big( \ved +\lambda - \vhd \eta_\bk \big)^2 - (\vpd \eta_\bk)^2 }. \label{ekb2}
\end{equation}

\subsubsection{Self-consistency equations}

The D-H binding saddle point solution can be obtained by solving the set of self-consistency equations derived from minimizing the energy with respect to the variables \{$\chi_d$, $\Delta_d$, $n_d$,
$p_{0\sigma}$, $\varepsilon$, $\lambda$, $\vhd$, $\vpd$, $\ved$ \}:
\begin{align}
&\vhd = 2tgp_{0\uparrow} p_{0\downarrow}\chi_f, \label{vchid} \\
&\vpd =g\chi_f \sum_\sigma \left( tp^2_{0\sigma} - g \Delta_d \vhf Y_\sigma \right), \\
&\ved = -\zeta g^2 \chi_f \vhf \sum_\sigma ( 1- p^2_{0\bar\sigma}) Y_\sigma, \\
&p_{0\sigma} = {2\zeta tg\chi_f (\chi_d p_{0\bar\sigma} +\Delta_d p_{0\sigma} ) \over \lambda -\lambda_\sigma -2\zeta g^2 \chi_f \vhf Y_{\bar\sigma} (1-n_d -p^2_{0\sigma}) }, \\
& p^2_{0\uparrow} -p^2_{0\downarrow} =n^f_{\uparrow} -n^f_{\downarrow}, \\
& 2n_d + p^2_{0\uparrow} + p^2_{0\downarrow} =1, \\
&\chi_d = d^2_0 + {1\over 2N\zeta} \sum_\bk {}^{'} \langle \eta_\bk d^\dagger_{A\bk} d_{B\bk} + h.c. \rangle, \label{chid} \\
&\Delta_d = d^2_0 + {1\over 2N\zeta} \sum_\bk {}^{'} \langle \eta^*_\bk d_{A\bk} e_{B\bar\bk} + h.c. \rangle,  \label{deltad} \\
&n_d = d^2_0 +{1\over 2N} \sum_{\alpha=\{ A, B\}} \sum_\bk {}^{'} \langle d_{\alpha\bk} d_{\alpha\bk} \rangle, \label{ndd}
\end{align}
where the fermion density $n^f_\sigma$ and hopping $\chi_f$ per spin is readily obtained from the fermion spectrum in Eq. (\ref{Efermion}).
It is instructive to examine the last three equations for the D/H hopping, the D-H binding, and the D/H density.
The closing of the boson gap $\Xi_d$ leads to a zero energy mode at $\bk=0$ whose occupation enables the single-boson condensate $d_0^2=e_0^2$.
This zero mode will be subsequently taken out of the momentum summations in Eqs. (\ref{chid}-\ref{ndd}).
Accordingly, the solutions to this set of self-consistency Eqs. (\ref{vchid}-\ref{ndd}) must be searched under two conditions: (i) assume $d_0=0$ and (ii) assume a nonzero $d_0$.
In the latter case, one more variable ($d_0$) is introduced together with one extra equation that ensures the existence of the zero energy mode,
\begin{equation}
\ved + \lambda = \zeta (|\vhd| +|\vpd|).
\end{equation}
If multiple solutions are found, the one with the lowest energy should be chosen as the ground state. In practice, we found only one solution at any given $U$.

\subsubsection{Electron spectral function and spectral density}

Once the saddle point solution is obtained, the spectral function of the physical electrons can be calculated from the one-particle Green's function $G_{\alpha\sigma}(\bk,\tau)
=-\langle T_\tau c_{\alpha \bk \sigma} (\tau) c^\dagger_{\alpha \bk \sigma} (0)\rangle$.
The detailed derivation of the spectral function and the integrated spectral function (ISF), i.e. the tunneling density of states,
\begin{equation}
N_{\alpha} (\omega) = -\text{Im} \int^\beta_0 d\tau e^{i\omega\tau} \sum_{\bk,\sigma} G_{\alpha\sigma} (\bk,\tau),
\end{equation}
are given in the Appendix. Note that since the spectral function involves convolutions of the $(d,e)$ boson normal and the anomalous (due to pairing) Green's functions with that of the $f_\sigma$-fermion,
the single-particle energy gap for the physical electron is the sum of the fermion and boson gaps  $\Xi=\Xi_d +\Xi_f$.
More importantly, the coherent QP excitations would only emerge with the D/H condensate that recombines the charge and spin degrees of freedom, and can be detected by the QP coherent peaks in $N(\omega)$.

\subsection{Ground state wavefunctions}

Before presenting the results on the honeycomb and the square lattice, it is instructive to discuss the possible phase structure in terms of the general form of the ground state wave function of the D-H binding saddle point. Since the Hilbert space is represented by those of the fermion and the slave bosons, the electron ground state wavefunction is a product of the ground state wave functions for the bosons and fermions,
\begin{align}
\Psi(\br_1\sigma_1,\dots, \br_N\sigma_N)=&\Psi_B(\br_1,\dots, \br_{N_d}; \br_1,\dots, \br_{N_e})
\nonumber \\
&\otimes \Psi_F(\br_1\sigma_1,\dots, \br_N\sigma_N).
\label{electronwf}
\end{align}
Here $\sigma_i, i=1,\dots,N$ labels the spins of $N$ electrons, while $N_d$ and $N_e$ are the number of doublons and holons, respectively.
From Eq. (\ref{Efermion}), it is clear that the fermion wavefunction is given by a Slater determinant, i.e. $\Psi_F =\Psi_{\rm Slater}(\{\br_i\sigma_i\})$ in both the PM and the AF ordered phases.
Comparing to the conventional wavefunction form for an interacting many-body electron system, $\Psi(\{\br_i\sigma_i\}) =\prod_{i<j}J(\br_i-\br_j)\Psi_{\rm Slater}(\{\br_i\sigma_i\})$, the variational Jastrow factor $J$ has been promoted to full-fledged boson wavefunctions, thus allowing possible new electronic phases.
The key physics in our theory is the boson inter-site correlations.
The corresponding boson ground state wavefunction in second quantized form is thus a direct product of single-boson condensates and the pairing of uncondensed doublons and holons \cite{rokhsar94}.
\begin{figure}
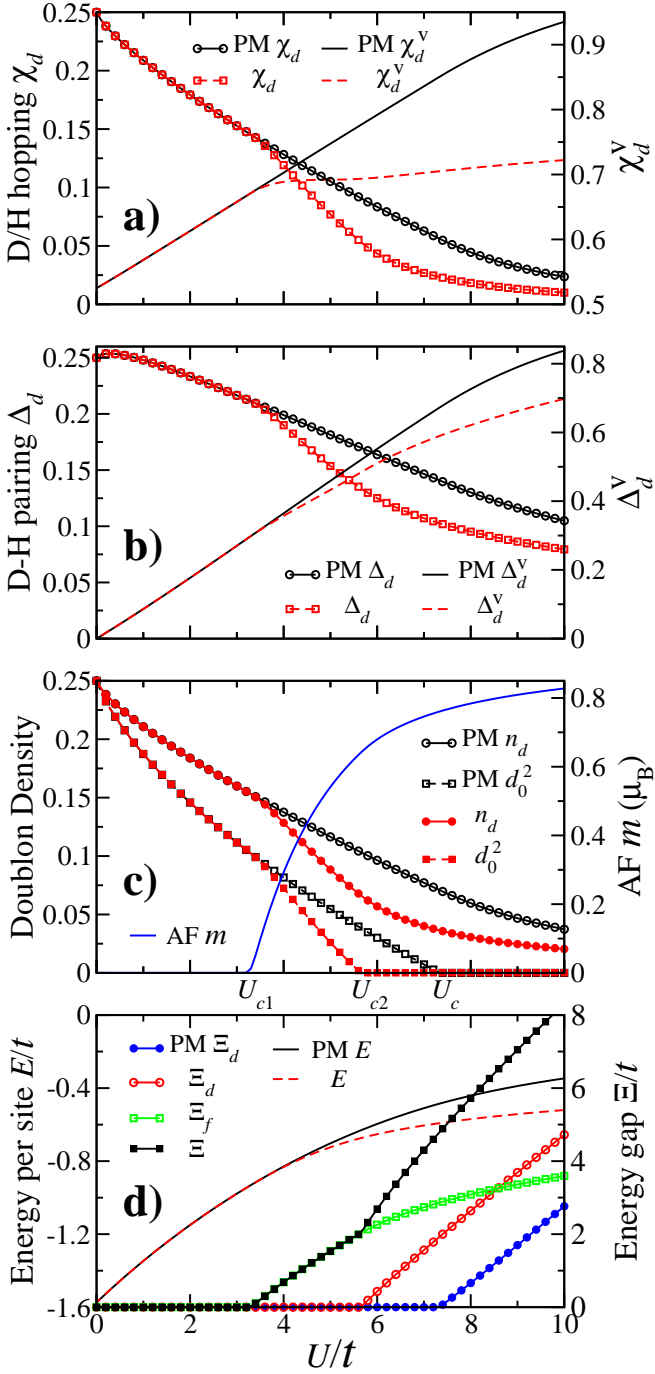

\begin{center}
\fig{3.4in}{honeycombDH.eps} \caption{(color online).
D-H binding saddle point solutions on the honeycomb lattice.
The Hubbard $U$ dependence of (a) variational parameters $\vhd$ and D/H hopping order parameter $\chi_d$; (b) variational parameter $\Delta^{\text{v}}_d$ and D-H pairing order parameter $\Delta_d$; (c) D/H density $n_d$, condensate density $d_0^2$, and AF staggered magnetization $m$; and (d) the ground state energy per site $E$, energy gaps in the boson sector $\Xi_d$, the fermion sector $\Xi_f$, and for the physical electrons $\Xi=\Xi_f +\Xi_d$. The corresponding results in the restricted PM phase are also shown.
In the restricted PM phase, the Mott transition (SM to Mott insulator) takes place at $U_c\simeq 7.3t$.
While the ground state undergoes two transitions: the magnetic transition (from SM to Slater AF insulator) at $U_{c1}\simeq 3.4t$ and the Mott transition (from Slater AF to AF$^*$ phase) at $U_{c2}\simeq 5.7t$.
}\label{honeycombDH}
\end{center}
\end{figure}
\begin{figure}
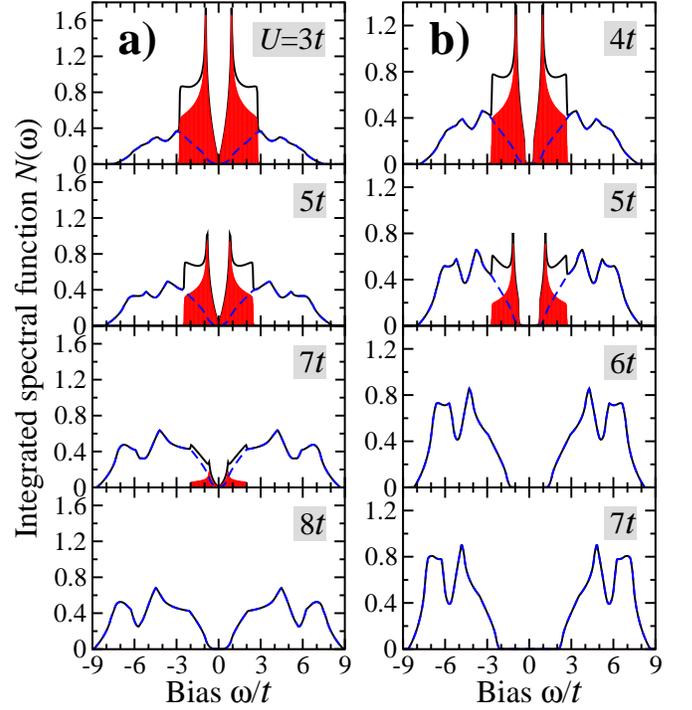

\begin{center}
\fig{3.4in}{honeycombDOS.eps} \caption{(color online). ISF with D-H binding on the honeycomb lattice. The coherent (shaded red areas), incoherent (blue solid lines), and the total (black solid lines) integrated spectral function at different values of Hubbard $U$ in (a) the restricted PM phase where the Mott transition is at $U_c\simeq7.3t$ and (b) the ground state where the AF to AF$^*$ transition is at $U_{c2}\simeq5.7t$.}
\label{honeycombDOS}
\end{center}
\end{figure}
On the square lattice, the boson wavefunction from the Hamiltonian in Eq.~(\ref{Eboson2}) thus takes the form
\begin{equation}
\vert\Psi_B\rangle=(e_0^\dagger)^{N_e^0}(d_0^\dagger)^{N_d^0} \vert0\rangle_B \otimes \prod_\bk \exp(-g_\bk d^\dagger_\bk e^\dagger_{\bar\bk}) \vert 0\rangle_B, \label{bosonwf}
\end{equation}
where $\vert0\rangle_B$ is the vacuum of the boson sector, and the pairing function $g_\bk=[\ved+\lambda-\vhd\eta_\bk -E^d(\bk)] /\vpd \eta^*_\bk$.
The boson wavefunction on the honeycomb lattice described by the Hamiltonian in Eq. (\ref{Eboson}) has a similar, slightly more complicated form due to the doubling of the unit cell.
In the first part of the boson wavefunction, $N_e^0$ and $N_d^0$ are the numbers of condensed holons and doublons that are determined by the condensate density $e_0^\dagger e_0$ and $d_0^\dagger d_0$.
This is the only part retained in the KR saddle point or the Gutzwiller approximation.
Together with $\Psi_F$, they describe the coherent QP part in the excitation spectrum of a correlated Fermi fluid.
The second part is due to D-H binding which, together with $\Psi_F$, describes the incoherent excitations and plays an integral part in the Mott transition.
So long as the condensate part is present, the charge and spin degrees of freedom combine such that the elementary excitations carry the quantum numbers of an electron and appear as QP poles in the single-particle Green's function.
Thus, the coexistence of the condensate and the binding parts heralds the coherent QP and the incoherent Hubbard excitations in Mott-Hubbard systems before the Mott transition.
This is the case in both the PM ($U<U_{c1}$) and the Slater AF phase ($U_{c1} < U < U_{c2}$) where $N_e^0=N_d^0\ne0$ in the condensed part of the boson wavefunction $\Psi_B$ while $\Psi_F$ changes from a PM to AF Slater determinant.
However, as we will show, when $U>U_{c2}$, the quantum fluctuations due to D-H binding destroy the single-particle condensate, i.e. $N_e^0=N_d^0=0$, as all doublons and holons are bound together, giving rise to a charge gap.
The boson wave function is given entirely by the second part in Eq.~(\ref{bosonwf}).
Interestingly, this boson wave function is just the wave function of a resonating valence bond (liquid) state, which in the present context can also be understood as that of an excitonic insulator, since the doublon and holon carry opposite charges. Because all the doublons are bound to the holons, the elementary excitations do not carry the quantum numbers of an electron and the entire single-particle excitations are incoherent as the charge and spin cannot recombine to form a coherent quasiparticle. Had this Mott transition taken place before the AF order, this insulating phase would be a spin liquid. However, as we will see that on the bipartite lattices without frustration, AF order happens before the Mott transition, i.e. $U_{c2} > U_{c1}$. We thus term the phase for $U>U_{c2}$ the AF* phase, which is indeed an example of spin-charge separation above one-dimention, albeit taking place inside the AF ordered phase.
Note that what distinguishes the AF* phase from the Slater AF insulator is the complete depletion of the single-particle condensates of the holons and doublons above $U_{c2}$ such that all doublons are bound to holons.

\subsection{Mott, Slater AF, and AF Mott transitions}

\subsubsection{Results on the honeycomb lattice}

The D-H binding saddle point solutions on the half-filled honeycomb lattice are summarized in Figs.~\ref{honeycombDH} and \ref{honeycombDOS}.
The variational parameters $\vhd$ and $\vpd$, the order parameter for the D/H hopping $\chi_d$ and D-H pairing $\Delta_d$ are plotted in Figs.~\ref{honeycombDH}a and \ref{honeycombDH}b as function of the Hubbard $U$.
At $U=0$, $\vpd=0$, thus all doublons and holons are single-particle condensed with $d_0^2 =e_0^2 =p_{0\sigma}^2 =1/4$ such that $\vhf=t$, recovering the noninteracting limit.
The SM phase remains stable at small $U$.
With increasing $U$, the doublon density decreases as shown in Fig.~\ref{honeycombDH}c.
Due to the increase in D-H binding, the D/H condensate $d_0$ decays faster than in the KR saddle point solution shown in Fig.~\ref{honeycombGW}a.
To study the Mott transition, we first restrict the solution to be in the PM phase by enforcing $p_{0\uparrow} =p_{0\downarrow}$, which amounts to suppressing possible magnetically ordered states.
As shown in Fig. \ref{honeycombDH}c, the Mott transition takes place at $U_c\simeq 7.3t$, which is considerably smaller than that of $12.6t$ for the BR transition (Fig.~\ref{honeycombGW}a).
The condensate $d_0$ vanishes and all doublons are bound with the holons in the Mott insulating phase for $U>U_c$, accompanied by the opening of a charge gap $\Xi_d$ that is linear in $U-U_c$ (Fig. \ref{honeycombDH}d).
The ISF of the physical electrons is shown in Fig. \ref{honeycombDOS}a.
Notice the transfer of the coherent QP weight to the incoherent part with increasing $U$ and the complete suppression of the coherent QPs in favor of two broad incoherent spectral features beyond the Mott transition that originate from the bosonic excitations $E^d_\pm (\bk)$ in Eq. (\ref{Eboson}) to be discussed later.
Since the $f_{i\sigma}$-fermion spinon remains gapless, the insulating phase is a gapless SL.
Thus, we find no evidence on the honeycomb lattice for the proposed gapped SL phase \cite{meng}.

Next, we allow magnetism and study the interplay between AF order, D-H binding, and the Mott transition in the ground state. Fig.~\ref{honeycombDH}c shows that the SM phase on the honeycomb lattice remains stable until a critical $U_{c1}\simeq3.4t$, where the staggered magnetization ($m$) onsets. We find that for $U_{c1}<U<U_{c2}$, where $U_{c2}\simeq5.7t$, although a single-particle gap $\Xi_f$ opens in the fermion sector (Fig. \ref{honeycombDH}d), the zero energy mode remains stable in the $d$-$e$ sector and continues to support the D/H condensate. Thus, the spin and charge continues to recombine in this regime and there are coherent excitations corresponding to the sharp QP peaks in the ISF shown in Fig. \ref{honeycombDOS}b at $U=4t$ and $5t$.
This phase is thus an Slater AF insulator whose wavefunction would overlap well with an AF Slater determinant.

Remarkably, a Mott transition in the presence of AF order takes place at $U_{c2}$.
For $U>U_{c2}$, an AF Mott phase (i.e., the AF$^*$ phase) emerges with the opening of the boson gap $\Xi_d\propto U-U_{c2}$ in the $d$-$e$ sector (Fig. \ref{honeycombDH}d) as the D/H condensate vanishes. Since all doublons are bound to holons, the charge and spin cannot recombine and the electrons are thus fractionalized in the AF$^*$ phase. A direct consequence for the lack of elemental excitations carrying the electron quantum number is that the lack of coherent QP peaks in an entirely incoherent excitation spectrum,
as can be seen from the broad ISF at $U>U_{c2}$ in Fig. \ref{honeycombDOS}b at $U=6t$ and $7t$.
Unlike in the Slater AF phase, the vanishing of the D/H condensate in the AF$^*$ phase,
enables the deconfinement the spin and charge degrees of freedom,
such that the ground wavefunction has no overlap with Slater determinant-like states.
The excitation energy gap for the physical electron, $\Xi=\Xi_f+\Xi_d$, exhibits a derivative discontinuity at $U_{c2}$ (Fig.~\ref{honeycombDH}d) due to the opening of the Mott gap $\Xi_d$ in the AF$^*$ phase.
However, the magnetization $m$ remains analytic across the AF$\to$AF$^*$ transition in Fig. \ref{honeycombDH}c,
which is a topological confinement-deconfinement transition associated with the Ising-like global $Z_2$ symmetry ($d_i\to-d_i$, $e_i\to-e_i$) that is broken in the Slater AF phase by the D/H condensate and restored in the AF$^*$ phase.

The continuous SM to AF transition at $U_{c1}\simeq3.4t$ compare well to the most recent QMC calculations on large system sizes by Sorella et. al. \cite{sorella} that finds the onset of AF order and a single-particle excitation gap at $U\simeq3.8t$. Since we have not included the inter-site spin fluctuations described by the dynamics of the $p_\sigma$-boson, our magnetic gap is somewhat larger than the QMC values, and we will not attempt quantitative comparisons to results obtained by other numerical methods such as cluster dynamical mean-field theory (CDMFT) calculations with continuous-time QMC (CTQMC) or exact diagonalization (ED) impurity solvers.
While the CTQMC-CDMFT \cite{wliu,hur12} is performed at relative high temperatures and not very suitable for extracting small energy gaps in the quantum states, the ED-CDMFT \cite{liebsch, zylu} as well as the ED-VCA (variational cluster approximation) \cite{jxli, seki} revealed spurious excitation gaps at very small $U$, before the emergence of AF order. This was viewed as supporting evidence for the proposed gapped SL phase \cite{meng}.
Recently, Hassan and S\'{e}n\'{e}chal \cite{hassan13} noticed that these ED-CDMFT and ED-VCA calculations use only a single bath orbital per cluster site which they argued is insufficient and leads to the artificial excitation energy gaps for all nonzero values of $U$. Their calculations with two bath orbitals connecting each cluster site show that the PM Mott transition and thus the SL phase is indeed preempted by a magnetic transition occurring at a lower value of $U$. Liebsch and Wu\cite{liebsch2} further pointed out that the spurious excitation gap at very small $U$ originates from the breaking of the translation symmetry in these cluster calculations.

\begin{figure}
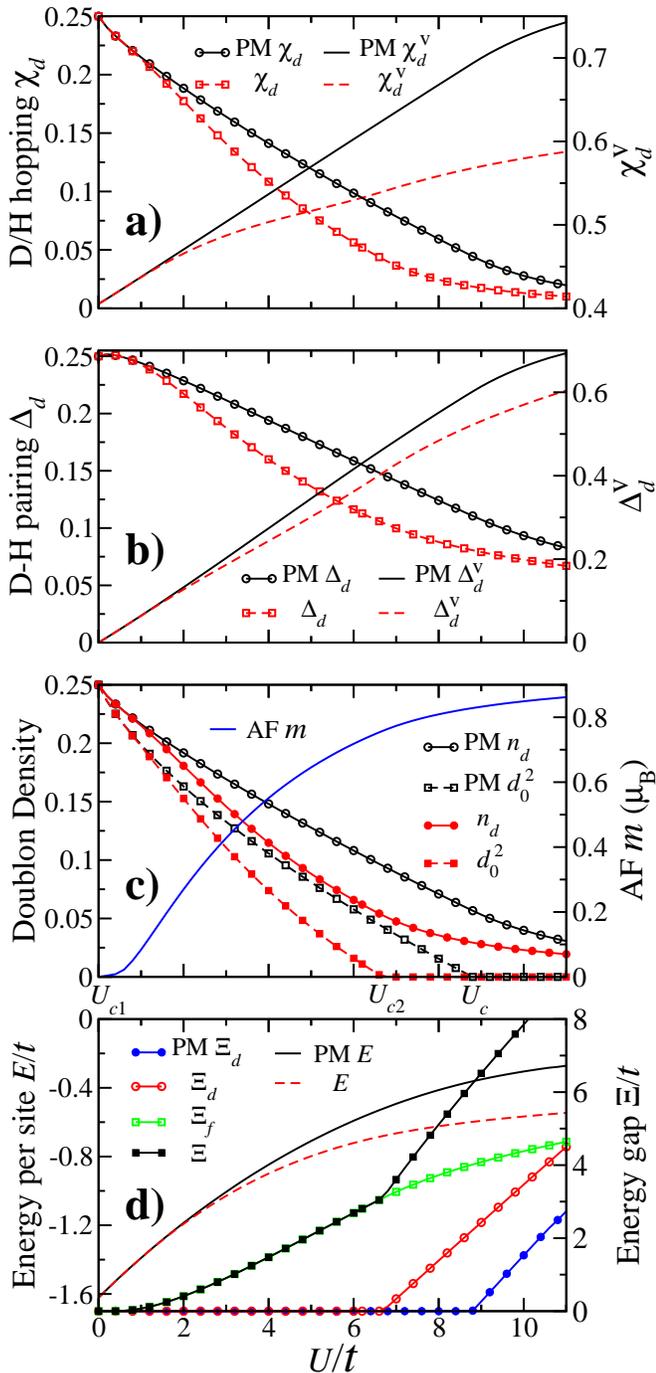

\begin{center}
\fig{3.4in}{squareDH.eps}
\vspace{-0.1cm}
\caption{(color online).
D-H binding saddle point solutions on the square lattice,
where $U_c\simeq 8.8t$, $U_{c1}=0$, and $U_{c2}\simeq 6.8t$.
The labels and notations follow those in Fig. \ref{honeycombDH}.}
\label{squareDH}
\end{center}
\end{figure}

\begin{figure}
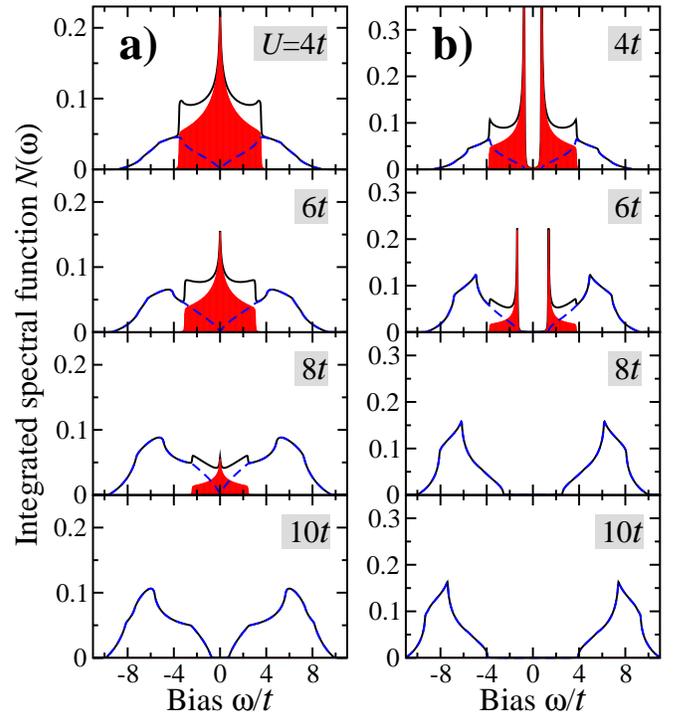

\begin{center}
\fig{3.4in}{squareDOS.eps} \caption{(color online).
ISF with D-H binding on the square lattice.
The labels and notations follow those in Fig. \ref{honeycombDOS}.}
\label{squareDOS}
\end{center}
\end{figure}

\subsubsection{Results on the square lattice}

Figs.~\ref{squareDH} and \ref{squareDOS} show the results obtained on the half-filled square lattice, which are qualitatively the same as those obtained on the honeycomb lattice. In the PM phase, the Mott transition is at $U_c\simeq8.8t$ in contrast to the Brinkman-Rice transition at $U_{\rm BR}\simeq13t$ without taking into account of D-H binding. Because the band structure leads to perfectly nested Fermi surface at half-filling, the PM metallic phase is unstable towards AF order for infinitesimal $U$ on the square lattice. The AF order therefore emerges at $U_{c1}=0$ with an exponentially small staggered magnetization, as shown in Fig.~\ref{squareDH}c. This Slater AF insulator with coherent QP excitations is stable until $U_{c2}\simeq6.8t$ where a transition into the AF* phase with the vanishing of holon/doublon condensate and the opening of the charge gap (Fig.~\ref{squareDH}c,d) and the disappearance of coherent QP peaks in favor of two broad incoherent features in the integrated spectral density Fig.~(\ref{squareDOS}). It is important to note that the Fermi level density of states, whose two limiting behaviors, vanishing or divergent, are presented by the unfrustrated honeycomb and square lattices respectively, while affecting the PM to Slater AF transition, does not play an essential role in determining the Mott transition from the Slater AF into the AF* phase. This is because the latter is tied to the opening of the charge gap in the boson sector near the doublon and holon band bottom above a finite $U_{c2}$, as can be seen from the boson dispersions shown in Fig.~\ref{dispB}.

To further explore the generality of these predictions, we have studied the case where the noninteracting band has a semicircular density of state $\rho(\omega)=(4/\pi W) \sqrt{1-(2\omega/W)^2}$ where $W$ is the bandwidth. We found an identical phase structure with a PM metal to Slater AF insulator at $U_{c1}\simeq0.1W$, followed by the Mott transition into that AF* phase at $U_{c2}\simeq W$. Note that although the semicircle density of states can be realized in the infinite-d unfrustrated Bethe lattice, these results should not be considered as obtained for the infinite-d Hubbard model, since taking the infinite-d limit would suppress all inter-site correlations, including the inter-site D-H binding considered here. Thus, in the infinite-d limit, we would only recover the KR saddle point solution, i.e. the Gutzwiller approximation which is exact for the Gutzwiller wave function approach in infinite dimensions \cite{GWinf-d}. In this sense, our approach can be viewed as going beyond the Gutzwiller approximation by including the inter-site correlations in physical dimensions.

\begin{figure}
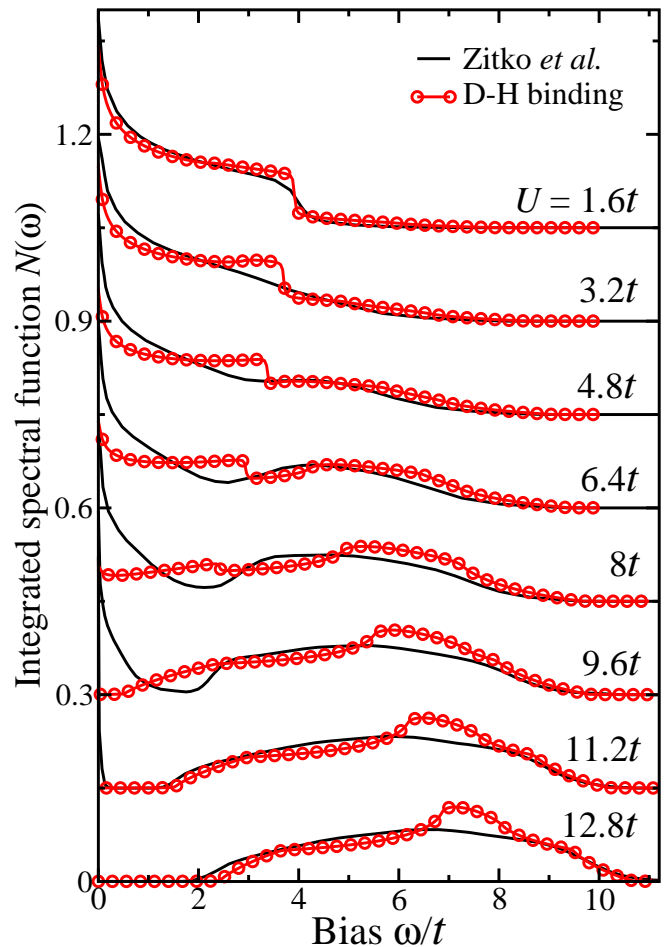

\begin{center}
\fig{3.4in}{square-dos.eps} \caption{(color online).
Local spectral function obtained from the D-H binding theory and the single-site DMFT by \v{Z}itko \textit{et al.}\cite{zitko09} on the paramagnetic square lattice at various $U$.
The curves are offset vertically for clarity.}
\label{square-dos}
\end{center}
\end{figure}

With this difference in mind, we proceed to compare in Fig.~\ref{square-dos} the local spectral function in the PM phase on the square lattice of our D-H binding theory with the results obtained from the single-site DMFT on the square lattice \cite{zitko09} at different Hubbard $U$. Overall, we find remarkable agreement in the incoherent part of the spectrum for all values of $U/t$ on both sides of the Mott transition. The most significant deviations of the results are in the low energy QP peaks around the Mott transition. The main cause of the latter can be traced to a particular property of the single-site DMFT formulation that the spectral density at $\omega=0$ is independent of $U$ in the infinite-d limit of the Hubbard model \cite{muller-hartmann}, which holds the QP peak at constant height until its width goes to zero at the Mott transition, seen from the DMFT data in  Fig.~\ref{square-dos}. The latter is no longer true in the presence of inter-site correlations beyond the infinite-d limit, as shown in the D-H binding results in Fig.~\ref{square-dos}, where the QP peaks, agreeing with the DFMT result for moderate $U$, are suppressed in both height and width and disappears completely at the Mott transition whose critical value is significantly reduced by intersite correlations \cite{park08}. We also find broad qualitative agreement with the local spectral density of cluster DMFT calculations that captures certain short-range correlations, although finding consistency in CDMFT results is difficult due to the different cluster embedding procedure (including cluster shapes and sizes) and the choice of the impurity solver. We find that it is particularly intriguing that in the CDMFT study of the PM phase by Park, et. al. \cite{park08}, the metallic phase has an ISF consistent with our result in the PM phase; whereas on the insulating side near the metal-insulator transition, the local spectral function displayed a small gap with very pronounced peaks at the gap edge that closely resemble our findings in the AF Slater insulator. Indeed, these peaks at the edge of the magnetic gap are clear hallmark of the coherence QP peaks characteristic of a Slater spin density wave insulator. We thus conjecture that the CDMFT findings of a small gap PM insulating state with pronounced gap-edge coherence peaks are due to the fluctuating or short-range Slater AF order, and with increasing $U$, a true Mott transition would emerge with the suppression of the QP peaks and the opening of the charge gap.

\subsubsection{Stability of the D-H binding saddle point}

Next we comment on the D-H binding saddle point stability with respect to gauge field fluctuations.
It is known that KR formulation introduces three $U(1)$ gauge fields\cite{rasul88} since the action is invariant under:
$e_i\to e_i e^{i\theta_i}$, $p_{i\sigma}\to p_{i\sigma} e^{i\phi_{i\sigma}}$, $d_i\to d_i e^{-i\theta_i+i\sum_\sigma \phi_{i\sigma}}$, $f_{i\sigma}\to f_{i\sigma} e^{i\theta_i -i\phi_{i\sigma}}$,
and $\lambda_i \to\lambda_i +{\dot \theta}_i$,$\lambda_{i\sigma}\to
\lambda_{i\sigma}+{\dot\theta}_i-{\dot\phi}_{i\sigma}$.
The $p_{i\sigma}$ condensate breaks two of the $U(1)$ symmetries and turns the gauge fields associated with $\phi_{i\sigma}$ massive by the Anderson-Higgs mechanism.
The remaining $U(1)$ symmetry is also broken in the SM and the AF phase by the D/H condensate,
making the $\theta_i$-gauge field massive.
In the AF$^*$ phase, it is the D-H pairing $\Delta_{ij}$ that breaks the $U(1)$ symmetry and the $\theta_i$-gauge field remains massive, as does its staggered component due to the D/H hopping fields $\chi_{ij}^{d,e}$.
The absence of gapless gauge field fluctuations supports the stability of the obtained phases.

\subsection{Nature of incoherent Mott-Hubbard excitations}

\begin{figure}
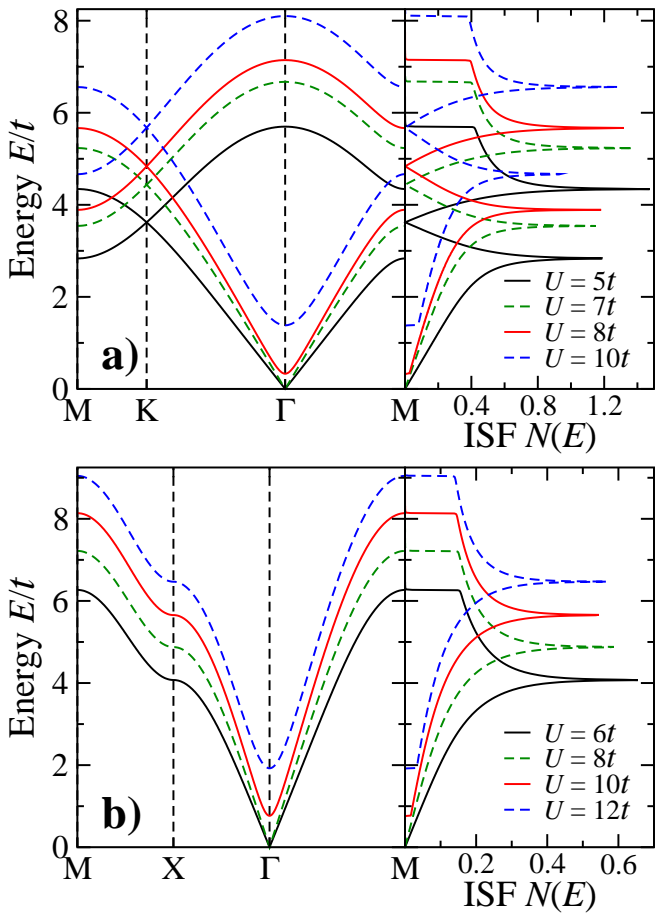

\begin{center}
\fig{3.4in}{dispB.eps} \caption{(color online). The energy spectrum along
high symmetry directions (left panel) and the ISF (right panel) of the
doublons and holons in the restricted PM phases at several Hubbard $U$
across the Mott transition on (a) the honeycomb lattice where $U_c\simeq 7.3t$ and (b) the square lattice where $U_c\simeq 8.8t$.
The single-particle condensate contribution  to the ISF (a delta function at zero energy) on the metallic side of the Mott transition is not shown.}
\label{dispB}
\end{center}
\end{figure}

It is enlightening to discuss the energy spectrum and the spectral function
of the doublons and holons in connection to the nature of the incoherent
Mott-Hubbard excitations in the local spectral function.
The dispersion of the holons and doublons in Eqs.~(\ref{ekb}) and (\ref{ekb2}) and the
corresponding density of states are shown in Fig.~\ref{dispB} in the PM
phases of the honeycomb and square lattice Hubbard models, respectively, at several values of $U$ across the Mott transition. Their behaviors are similar in the AF and AF$^*$ phases.

\noindent{\em Honeycomb Lattice}: At a fixed $U$, the boson spectrum shows two doubly-degenerate dispersive
branches given in Eq.~(\ref{ekb}). There are several noteworthy features.
(i) Both dispersive branches are flat near the $M$ point of the hexagonal Brillouin zone, leading to the two Van Hove singularities in the boson ISF plotted in the right panel of Fig.~\ref{dispB}a.
(ii) The two branches cross and produce the Dirac cone at the high symmetry $K$ point at a finite energy that increases with increasing $U$, leading to the V-shaped density of states.
Remarkably, (i) and (ii) combine to form a Dirac-cone like dispersion that is similar and can be regarded as a ``ghost'' band of the bare electron dispersion carried by the excitations of the D/H complex.
This property was pointed out in the systematic large-$N$ expansion study of the $t$-$J$ model for doped Mott insulators \cite{largeN}.
It is remarkable that the ghost Dirac-cone feature manifests itself in the broad peak-dip-peak structure in incoherent part of the ISF for the physical electrons shown in Fig.~\ref{honeycombDOS}a, which can now be identified as the D-H excitations.
(iii) The low energy properties of the boson dispersion near the $\Gamma$-point is also intriguing.
For $U < U_c$, i.e. on the metallic side of the Mott transition, the lower energy branch is gapless, i.e. $E^d_- (\bk)=0$ and disperses linearly away from the $\Gamma$ point.
The existence of the zero-energy mode, together with the vanishing of the ISF $N(E)$, enables the finite-temperature D/H condensation in a two-dimensional system such that $d_0\ne0$ at zero temperature.
In contrast, for $U>U_c$, or $U>U_{c2}$ when magnetism is allowed, a charge gap $\Xi_d\ne0$ opens up at the $\Gamma$ point, indicating the emergence of the Mott insulating or the AF$^*$ phase with complete suppression of the D/H condensate. Note also that the gapped $E^d_- (\bk)$ is parabolic near $\Gamma$, giving rise to a finite ISF $N(E)$ at the band bottom.

\noindent{\em Square Lattice}: The above findings on the connection between the incoherent Mott-Hubbard excitations and the doublon-holon spectrum applies in a straightforward manner to the square lattice case as well. In contrast to the intrinsic two-sublattice structure of the honeycomb lattice, the boson spectrum on the square lattice has only one branch given in Eq.~(\ref{ekb2}), which is shown in Fig.~\ref{dispB}b. The corresponding density of states has a single Van Hove peak tied to the dispersion near the X point. Similar to the honeycomb lattice case, this branch of D-H excitations manifest itself in the single broad peak on the particle and the hole side of the electron local spectral function shown in Fig.~\ref{squareDOS}.

The change in the bosonic dispersion across the AF Mott transition, i.e. from the Slater AF insulator to the AF* phase, is qualitatively the same as those displayed in Fig.~\ref{dispB} for the PM Mott transition. Namely, the boson spectrum is gapless with linear dispersion supported by the D/H single-particle condensate on the Slater AF side and develops an energy gap, when all holons are bound to doublons, above which a quadratic dispersion is found for the Bogoliubov quasiparticles. The linear dispersion in the PM phase is a bosonic representation of the collective zero-sound excitations in the Landau Fermi liquid. In the Slater AF phase, despite the opening of a single-particle magnetic gap, the absence of a charge gap is reflected in the existence of gapless collective excitations represented by the bosons. It is only after entering the AF* phase, the charge excitations are gapped out, leaving only the spin-waves as the low energy excitations inside the single-particle energy gap.

\section{Conclusions}

In summary, we have shown that the binding between doublons and holons plays an essential role in describing the incoherent excitations and the Mott transition
in strongly correlated Mott-Hubbard systems.
For the honeycomb lattice Hubbard model, we showed that the SM to AF Slater insulator transition is followed by
a Mott transition into a fractionalized AF$^*$ phase with increasing $U$.
Interestingly, a different AF$^*$ phase of a fractionalized antiferromagnet was proposed in the effective Z$_2$ gauge theory description of doped Mott
insulators in the projected ($U=\infty$) Hilbert space where spinons are paired into a N\'eel state and doublons are absent \cite{fishersenthil}.
In contrast, the incoherent charge excitations through D-H binding is essential in the AF$^*$ phase proposed here, which is more inline with the importance of doublons in describing Mottness emphasized recently \cite{phillips10}.

The most practical way to test our predictions is to measure the energy gap for single-particle excitations using spectroscopic probes such as ARPES and STM. Our theory shows that with increasing $U/W$ (which can be varied experimentally by applying pressure or isoelectronic chemical substitution), the system goes from a gapless PM state to an AF insulator where the single-particle gap is controlled by the magnetic gap, followed by a transition to the AF* phase where a charge gap opens and adds on top of the magnetic gap. Thus, there is a singularity (kink) in the evolution of the gap as a function of $U/W$. Perhaps even more directly, the single-particle spectral function as measure by ARPES shows well defined QP peaks above the magnetic gap in the Slater AF phase, but would exhibit no coherent excitations in the AF* phase. Such an AF$^*$ phase on the square lattice may have been observed in the parent AF insulating compound of the high-$T_c$ cuprates by ARPES experiments \cite{arpesundoped}, which find no coherent QP excitations at all energies.

As a concrete example, we propose to revisit the time-honored Mott-Hubbard system, i.e. the transition metal oxide V$_2$O$_3$, under chemical pressure achieved by Cr or Ti substitutions. In this case, a finite temperature Mott metal-insulator transition above the low-temperature AF insulating ground state has been well established as a function of chemical substitution \cite{kuwamoto,mcwhan71,mcwhan73}.  Our theory predicts that hidden inside the AF insulating ground state is a transition from a Slater AF to the AF* phase. Moreover, melting the AF order in the Slater AF insulator would result in the metallic state, whereas melting the AF* phase at higher Cr substitutions gives rise to the Mott insulator at finite temperatures. Performing the experiments described above in these materials would either provide support or disprove our theory.

We thank F. Wang and Y. Yu for useful discussions.
This work is supported by DOE grants DE-FG02-99ER45747 and DE-SC0002554, and the Thousand Youth Talents Plan of China (SZ). ZW thanks Aspen Center for Physics for hospitality.

\appendix
\section{ISF of physical electrons}
The ISF, which equals the tunneling density of states (DOS), for the
physical electrons is given by
\begin{equation}
N_{\alpha} (\omega)= - \sum_{\bk,\sigma} \text{Im} \int^\beta_0 d\tau e^{i
\omega\tau} G_{\alpha\sigma}(\bk, \tau), \nonumber
\end{equation}
where retarded single-particle Green's function \cite{manhan}
\begin{equation}
G_{\alpha\sigma}(\bk,\tau)= -\langle T_\tau c_{\alpha \bk \sigma} (\tau)
c^\dagger_{\alpha \bk \sigma} (0)\rangle, \nonumber
\end{equation}
with $\alpha$ the sublattice index.
In the KR slave boson formulation \cite{kr}, the electron operator is composed of
$c_{i\sigma}={\hat L}_{i\sigma}(e_i^\dagger p_{i\sigma}
+p_{i{\bar\sigma}}^\dagger d_i){\hat R}_{i\sigma}f_{i\sigma}$.
Within our saddle point solution, $\hat{L}_{i\sigma}$ and
$\hat{R}_{i\sigma}$ are approximated by their saddle point average for the local Green's functions.
The electron operator in momentum space is thus given by,
\begin{equation}
c_{\alpha \bk \sigma} = r_{\alpha\sigma} \sum_{\bq,\bq'} \big(
e^\dagger_{\alpha\bar\bq} p_{\alpha\bq'\sigma} +p^\dagger _{\alpha
\bar \bq' \bar\sigma} d_\bq \big) f_{\alpha,\bk-\bq-\bq', \sigma},
\nonumber
\end{equation}
with the normalization factor
\begin{equation}
r_{\alpha\sigma} = \langle \hat{L}_{\alpha\sigma} \hat{R}_{\alpha\sigma}
\rangle = \big[ (1-n_d -p^2_{\alpha 0 \sigma}) (1-n_e -p^2_{\alpha 0
\bar \sigma})\big]^{-1/2}. \nonumber
\end{equation}
Therefore, the electron Green's function
\begin{equation}
G_{\alpha\sigma}(\bk,\tau) =r^2_{\alpha\sigma} \sum_{\bq, \bq'}
\Lambda_{\alpha \sigma}(\bq,\bq',\tau) G^f_{\alpha \sigma} (\bk -\bq
-\bq',\tau), \nonumber
\end{equation}
where $G^f_{\alpha\sigma}(\bk,\tau)= -\langle T_\tau f_{\alpha \bk \sigma}
(\tau) f^\dagger_{\alpha \bk \sigma} (0)\rangle$ is the $f_\sigma$-fermion
Green's function which can be computed easily in terms of the fermionic QPs defined in Eq. (\ref{Efermion}), and $\Lambda$ involves the normal and
anomalous (due to
pairing) Green's functions of the bosons
\begin{align}
\Lambda_{\alpha\sigma}(\bq,\bq',\tau) = &
\langle T_\tau e^\dagger_{\alpha \bar\bq} (\tau) e_{\alpha \bar \bq} (0)
\rangle \langle T_\tau p_{\alpha \bq' \sigma} (\tau) p^\dagger_{\alpha
\bq'\sigma} (0) \rangle  \nonumber \\
+& \langle T_\tau d_{\alpha \bq} (\tau) d^\dagger_{\alpha \bq} (0) \rangle
\langle T_\tau p^\dagger_{\alpha \bar\bq' \bar\sigma} (\tau) p_{\alpha
\bar\bq' \bar\sigma} (0) \rangle \nonumber \\
+ &\langle T_\tau e^\dagger_{\alpha\bar\bq} (\tau) d^\dagger_{\alpha\bq}
(0) \rangle \langle T_\tau p_{\alpha \bq' \sigma} (\tau) p_{\alpha
\bar\bq'
\bar\sigma} (0) \rangle \nonumber \\
+& \langle T_\tau d_{\alpha\bq} (\tau) e_{\alpha \bar \bq} (0) \rangle
\langle T_\tau p^\dagger_{\alpha \bar\bq' \bar\sigma} (\tau)
p^\dagger_{\alpha \bq' \sigma} (0) \rangle. \nonumber
\end{align}
The ISF of the physical electrons becomes
\begin{equation}
N_{\alpha} (\omega) = - \sum_{\bk,\sigma} r^2_{\alpha \sigma} \text{Im}
\int^\beta_0 d\tau e^{i \omega \tau} \Lambda_{\alpha \sigma} (\tau)
G^f_{\alpha\sigma}(\bk, \tau) \label{isf2}
\end{equation}
where $\Lambda_{\alpha \sigma} (\tau) = \sum_{\bq,\bq'}
\Lambda_{\alpha\sigma}(\bq,\bq',\tau)$.

It is instructive to write each boson operator as the sum of the condensate
and fluctuations: $b^{(\dagger)}_\bk = b_0 \delta_\bk +
\tilde{b}^{(\dagger)}_\bk$, where $b$ stands for the $(d,e,p_\sigma)$
bosons.
Although this is not necessary, doing so facilitates well the following
discussions of the coherent and incoherent contributions to the electron
spectral function.
Note that the fluctuations $\tilde{b}^{(\dagger)}_\bk$ are boson operators,
obeying boson commutation relations and the energy spectrum discussed
above.
Thus, the normal and anomalous boson Green's functions can be written as
\begin{equation}
\langle T_\tau b^{(\dagger)}_\bk(\tau) b'^{(\dagger)}_\bq (0)\rangle = b_0
b'_0 \delta_\bk \delta_\bq + \langle T_\tau
\tilde{b}^{(\dagger)}_\bk(\tau)
\tilde{b}'^{(\dagger)}_\bq (0)\rangle. \nonumber
\end{equation}
Decomposing the condensate and fluctuation contributions this way and
keeping the leading order fluctuations involving a single boson Green's
function, we have
\begin{equation}
\Lambda_{\alpha\sigma}(\tau)= \Lambda^\text{cond} _{\alpha \sigma} (\tau)
+ \Lambda^\text{fluc} _{\alpha\sigma} (\tau),
\end{equation}
where the condensate part
\begin{equation}
\Lambda^\text{cond}_{\alpha\sigma} (\tau) = d^2_0 (p_{0\uparrow}
+p_{0\downarrow})^2,\label{Lcond}
\end{equation}
and the fluctuation part
\begin{align}
&\Lambda^\text{fluc}_{\alpha\sigma}(\tau)=p^2_{\alpha 0\sigma} \sum_{\bq}
\langle T_\tau \te^\dagger_{\alpha \bar\bq} (\tau) \te_{\alpha \bar \bq}
(0) \rangle  \label{Lfluc} \\
&\hspace{1.2cm}+p^2_{\alpha 0\bar\sigma} \sum_{\bq} \langle T_\tau
\td_{\alpha \bq} (\tau) \td^\dagger_{\alpha \bq} (0) \rangle \nonumber \\
&\hspace{0.5cm}+d^2_0 \sum_\bq \Big[ \langle T_\tau \tp_{\alpha \bq \sigma}
(\tau) \tp^\dagger_{\alpha \bq \sigma} (0) \rangle + \langle T_\tau
\tp^\dagger_{\alpha \bar\bq \bar \sigma} (\tau)
\tp_{\alpha \bar\bq \bar\sigma} (0) \rangle \Big] \nonumber \\
&\hspace{0.5cm} +p_{0\uparrow} p_{0\downarrow} \sum_{\bq} \Big[ \langle
T_\tau \te^\dagger_{\alpha\bar\bq} (\tau) \td^\dagger_{\alpha\bq} (0)
\rangle
+\langle T_\tau \td_{\alpha\bq} (\tau) \te_{\alpha \bar \bq} (0) \rangle
\Big]. \nonumber
\end{align}
Correspondingly, the ISF in Eq.~(\ref{isf2}) can be written as
\begin{equation}
N_{\alpha} (\omega) = N^\text{coh}_{\alpha} (\omega)
+N^\text{incoh}_{\alpha} (\omega), \label{isf3}
\end{equation}
with
\begin{align}
&N^\text{coh(incoh)}_{\alpha} (\omega)= \label{isf4} \\
&\hspace{0.5cm}-\sum_{\bk,\sigma} r^2_{\alpha \sigma} \text{Im}
\int^\beta_0 d\tau e^{i \omega \tau}
\Lambda^\text{cond(fluc)}_{\alpha \sigma}(\tau) G^f_{\alpha\sigma} (\bk,
\tau). \nonumber
\end{align}
The coherent part of the ISF comes from the single-boson condensates that
recombine the charge and spin degrees of freedom, leading to coherent
QP excitations associated with the coherence peaks in the ISF.
Beyond the Mott transition, the condensate of doublons and holons vanishes,
and the coherent ISF is completely suppressed.

In the incoherent part of the ISF defined in Eq. (\ref{isf4}), the
convolution of the boson and fermion Green's functions gives broad spectral
features. Since the $p_\sigma$ bosons are fully condensed and their
fluctuations were ignored for simplicity within our D-H saddle point
solution, the question arises as to how to evaluation the corresponding
Green's functions in Eq.~(\ref{Lfluc}).
Notice that at the saddle point level, the local constraint in
Eq.~(\ref{constraint:1}) is only satisfied on average, i.e. $\langle \hat
Q_{i\alpha}\rangle=0$.
When fluctuations are considered, a consistent condition imposed by the
constraint is $\langle T_\tau \hat Q_\alpha(\tau)\hat Q_\alpha(0)\rangle=0$
where $\hat Q_\alpha=(1/N)\sum_{i\in\alpha}\hat Q_{i\alpha}$ with $N$ the
number of $\alpha$-sublattice sites. Evaluating the latter to the leading
order in the boson correlations, one gets the following relation
\begin{align}
&{p^2_{0\uparrow} +p^2_{0\downarrow} \over 2} \sum_\bk \Big[ \langle T_\tau
\tp^\dagger_{\alpha \bar\bk\downarrow} (\tau)
\tp_{\alpha \bar \bk\downarrow} (0) \rangle + \langle T_\tau \tp_{\alpha
\bk\uparrow} (\tau) \tp^\dagger_{\alpha \bk \uparrow} (0) \rangle \Big]
\nonumber \\
&+d^2_0 \sum_\bk \Big[ \langle T_\tau \te^\dagger_{\alpha \bar\bk}
(\tau) \te_{\alpha \bar \bk} (0) \rangle + \langle T_\tau
\td_{\alpha \bk} (\tau) \td^\dagger_{\alpha \bk} (0) \rangle \nonumber \\
&\hspace{1.45cm}+ \langle T_\tau \te^\dagger_{\alpha \bar\bk} (\tau)
\td^\dagger_{\alpha \bk} (0) \rangle + \langle T_\tau \td_{\alpha
\bk} (\tau) \te_{\alpha \bar \bk} (0) \rangle \Big] =0.
\nonumber
\end{align}
As a result, the Green's function of the $\tp_\sigma$ boson in
$\Lambda_{\alpha\sigma}^{\text{incoh}}$ given in Eq.~(\ref{Lfluc}) can be
expressed in terms of those of the $\td$-$\te$ bosons; leading to
\begin{align}
&\Lambda^\text{fluc}_{\alpha \sigma} (\tau) = \sum_\bq \Big\{
\rho^e_{\alpha\sigma} \langle T_\tau \te^\dagger_{\alpha
\bar\bq} (\tau) \te_{\alpha \bar \bq} (0) \rangle \label{Lfinal} \\
&\hspace{2.3cm}+\rho^d_{\alpha\sigma} \langle T_\tau \td_{\alpha \bq}
(\tau)
\td^\dagger_{\alpha \bq} (0) \rangle \nonumber \\
&\hspace{1.3cm}+ \rho^{de}_{\alpha\sigma}
\big[ \langle T_\tau \te^\dagger_{\alpha \bar\bq} (\tau)
\td^\dagger_{\alpha \bq} (0) \rangle + \langle T_\tau \td_{\alpha
\bq} (\tau) \te_{\alpha \bar \bq} (0) \rangle \big] \Big\},
\nonumber
\end{align}
where
\begin{align}
&\rho^e_{\alpha\sigma}= p^2_{\alpha 0\sigma} -\bar{p}^2_0,\quad
\rho^d_{\alpha\sigma}= p^2_{\alpha 0\bar \sigma} -\bar{p}^2_0, \nonumber
\\
&\rho^{de}_{\alpha\sigma}= p_{0\uparrow} p_{0\downarrow} -\bar{p}^2_0;
\quad \text{with}\quad \bar{p}^2_0 = 2d^4_0/( p^2_{0\uparrow}
+p^2_{0\downarrow}). \nonumber
\end{align}
The normal and anomalous Green's function of the fluctuating doublons and
holons involved in Eq.~(\ref{Lfinal}) can be evaluated using bosonic QPs
defined in Eq. (\ref{Eboson}).
It is thus ready to compute the incoherent ISF in Eq. (\ref{isf4}).
Remarkably, at $U=0$, $p_{0\sigma} = d_0 =e_0=1/2$, thus
$\rho_{\alpha\sigma}^e =\rho_{\alpha\sigma}^d =\rho_{\alpha\sigma}^{de}
=\rho_{\alpha\sigma}^{ed}=0$ in Eq.~(\ref{Lfinal}) and the incoherent
spectral function is therefore completely suppressed, recovering the
noninteracting limit.


\begin{thebibliography}{99}
\bibitem{normalorder}
Because of the properties of the projection-operator, the normal ordering
of the square roots in the normalization factors is not an issue
\cite{kr}.
\end{thebibliography}
\end{document}